\documentclass[twocolumn,
preprintnumbers,prl,
superscriptaddress
]{revtex4-1}

\usepackage[height=8.85in,width=6.45in]{geometry}

\usepackage[utf8]{inputenc}
\usepackage[T1]{fontenc}
\usepackage{amsmath}
\usepackage{amssymb}
\usepackage{mathtools}
\allowdisplaybreaks

\usepackage{slashed}
\usepackage{braket}
\usepackage[usenames,dvipsnames,svgnames,table]{xcolor}
\usepackage[colorlinks,citecolor=DarkGreen,linkcolor=FireBrick,urlcolor=FireBrick,linktocpage,breaklinks=true]{hyperref}
\usepackage{graphicx}
\usepackage{tikz}
\usepackage{times}
\usepackage[scaled]{couriers}

\usepackage{bm}

\usepackage{xcolor}
\usepackage{mdframed}

\renewenvironment{figure}[1][]{
  \begin{originalfigure}[#1]
    \begin{mdframed}[linecolor=black!0,backgroundcolor=black!1]
}{
    \end{mdframed}
  \end{originalfigure}
}

\def\Nequals#1{$\mathcal{N}{=}#1$}
\def\bC{\mathbb{C}}

\def\bZ{\mathbb{Z}}

\def\sA{\mathsf{A}}
\def\sD{\mathsf{D}}
\def\sF{\mathsf{F}}
\def\stF{\tilde{\mathsf{F}}}
\def\Kitaev{\mathsf{Kitaev}}
\newenvironment{modified}{\bgroup}{\egroup}
\def\modi#1{#1}
\def\ii{\mathrm{i}}

\usepackage{colortbl}
\definecolor{llightyellow}{rgb}{1.0, 0.95, 0.7}
\definecolor{llightblue}{rgb}{0.7, 0.9, 1.0}
\definecolor{llightpink}{rgb}{1.0, 0.85, 0.95}
\definecolor{llightgreen}{rgb}{0.7, 1.0, 0.4}
\colorlet{lightyellow}{llightyellow!50!white}
\colorlet{lightblue}{llightblue!50!white}
\colorlet{lightgreen}{llightgreen!50!white}
\colorlet{lightpink}{llightpink!50!white}

\usepackage{listings}
\begin{document}

\preprint{IPMU-20-0008, RUP-20-5}
\title{Fermionic minimal models}

\author{Chang-Tse Hsieh}
\affiliation{Kavli Institute for the Physics and Mathematics of the Universe (WPI), \\
 University of Tokyo,  Kashiwa, Chiba 277-8583, Japan}
\affiliation{Institute for Solid State Physics, University of Tokyo, Kashiwa, Chiba 277-8581, Japan}
\author{Yu Nakayama}
\affiliation{Department of Physics, Rikkyo University, Toshima, Tokyo 171-8501, Japan}
\author{Yuji Tachikawa}
\affiliation{Kavli Institute for the Physics and Mathematics of the Universe (WPI), \\
 University of Tokyo,  Kashiwa, Chiba 277-8583, Japan}

\begin{abstract}
We show that 
there is a fermionic minimal model, i.e.~a 1+1d conformal field theory which contains operators of half-integral spins in its spectrum, for each $c=1-6/m(m+1)$, $m\ge 3$.
This generalizes the Majorana fermion for $c=1/2$, $m=3$ and the smallest $\mathcal{N}{=}1$ supersymmetric minimal model for $c=7/10$, $m=4$.
We provide explicit Hamiltonians on Majorana chains realizing these fermionic minimal models.
\end{abstract}

\maketitle
\section{Introduction and summary}
The classification of the unitary minimal models of conformal field theory in 1+1 dimensions \cite{Belavin:1984aa, Friedan:1984aa,Cardy:1986aa, Kato:1987td,Cappelli:1986hf,Cappelli:1987xt} is one of the triumphs of theoretical physics in the late 20th century.
It was a milestone in our understanding of universality \modi{of critical phenomena in certain 2d classical statistical models}  \cite{Andrews:1984af, Huse:1984mn, Pasquier:1986jc} and 1+1d quantum systems \cite{Fendley:2006aa, Feiguin:2006ydp}.

As is well known, the central charge is of the form $c=1-6/m(m+1)$ for an integer $m\ge 3$.
The simplest case $m=3$ is the critical Ising model with $c=1/2$
and the next case $m=4$ is the tricritical Ising model with $c=7/10$.
Starting from $m=5$, there are at least two distinct models, called the A-type (or the diagonal) modular invariant and the D-type modular invariant;
for $m=5$, they are the tetracritical Ising model and the critical 3-state Potts model, respectively. 
Finally, there are exceptions when $m=11,12,17,18,29$ and $30$. 
The operators of these models have integer spins. In this sense they can all be called bosonic.

Let us recall that the critical Ising model can be transformed to a free massless Majorana fermion via the Jordan-Wigner transformation \cite{SCHULTZ:1964aa}. 
They are almost the same, so much so that careful distinctions were not routinely made in the old literature. 
We stress that they are distinct: The theory of Majorana fermion has an operator of spin 1/2, while the Ising model does not. 
Similarly, the unitary minimal models with \Nequals1 supersymmetry were classified, and the smallest nontrivial example has the central charge $c=7/10$, the same as the tricritical Ising model \cite{Friedan:1984aa,Friedan:1985aa}. It is also known that this supersymmetric minimal model is obtained from the tricritical Ising model by the Jordan-Wigner transformation and can appear in a strongly interacting Majorana chain  \cite{Rahmani:2015aa}.

We can summarize these old observations as saying that we have fermionic minimal models when $m=3$ and $4$. 
It is then a natural question to ask whether there are fermionic minimal models  with higher $m$.
The purpose of this Letter is to answer this question positively. 

\begin{modified}
The existence of fermionic minimal models as 1+1d theories in the continuum should not really come as a surprise, although it was not widely appreciated \footnote{%
The authors learned very recently that such theories were found in the late 1980s \cite{Petkova:1988cy,Furlan:1989ra}.
They thank Prof.~V. Petkova for information.
}.
This is because  there is a general method developed a few years ago \cite{YTCernLect,Karch:2019lnn} 
which allows us to turn any 1+1d bosonic model with non-anomalous $\bZ_2$ symmetry into a fermionic model,
and the bosonic minimal models have such a $\bZ_2$ symmetry.
The method, however, is quite abstract.
The main result of this Letter then is to make this construction more concrete by providing\end{modified}
explicit lattice realizations of fermionic minimal models by presenting a systematic construction of Majorana chains from quantum spin chains which give rise to bosonic minimal models at criticality.


\section{Analysis in the continuum}\label{sec:continuum}

\subsection{General analysis}\label{sec:continuum-general}
We will review the argument of \cite{YTCernLect,Karch:2019lnn},
which allows us to turn a 1+1 dimensional bosonic theory with non-anomalous $\bZ_2$ symmetry into a fermionic theory.
This method is  a simplified version of the ideas developed in 2+1 dimensions \cite{Gaiotto:2015zta,Bhardwaj:2016clt},
\begin{modified}
and can be considered as a variant of orbifolding by the $\bZ_2$ symmetry.
As such, we are going to recall the ordinary $\bZ_2$ orbifolding procedure first,
and then discuss the fermionization procedure.
\end{modified}

Let us consider a 1+1d quantum field theory $\sA$ with a non-anomalous $\bZ_2$ symmetry.
We would like to study the Hilbert space of states on $S^1$,
which can be either untwisted or twisted, depending on whether we introduce a twist by the $\bZ_2$ symmetry around the spatial $S^1$.
The untwisted and the twisted states can then each be decomposed into states even and odd under the $\bZ_2$ symmetry. 
We present  this decomposition in Table~\ref{states},
where $S$, $T$, $U$ and $V$ are generic symbols for states in the respective sectors.

\begin{table*}
\begin{tabular}{c|cc}
$\sA$ & \text{untwisted} & \text{twisted} \\
\hline
\text{even} & \cellcolor{lightblue} $S$ & \cellcolor{lightpink} $U$ \\
\text{odd} & \cellcolor{lightgreen} $T$ &  \cellcolor{lightyellow} $V$
\end{tabular}
\quad
\begin{tabular}{c|cc}
$\sD$ & \text{untwisted} & \text{twisted} \\
\hline
\text{even} & \cellcolor{lightblue} $S$ & \cellcolor{lightgreen} $T$  \\
\text{odd} & \cellcolor{lightpink} $U$&  \cellcolor{lightyellow} $V$
\end{tabular}
\quad
\begin{tabular}{c|cc}
$\sF$ & \text{antiperiodic} & \text{periodic} \\
\hline
\text{bosonic} & \cellcolor{lightblue} $S$ & \cellcolor{lightpink} $U$ \\
\text{fermionic} &  \cellcolor{lightyellow} $V$ &  \cellcolor{lightgreen} $T$
\end{tabular}
\quad
\begin{tabular}{c|cc}
$\stF$ & \text{antiperiodic} & \text{periodic} \\
\hline
\text{bosonic} & \cellcolor{lightblue} $S$ & \cellcolor{lightgreen} $T$ \\
\text{fermionic} &  \cellcolor{lightyellow} $V$ &  \cellcolor{lightpink} $U$
\end{tabular}
\caption{States of the models $\sA$, $\sD$, $\sF$ and $\stF$. \label{states}}
\end{table*}

Let us consider the theory $\sD$ obtained by taking the orbifold, or performing the gauging, by this $\bZ_2$ symmetry. 
The untwisted sector of the theory $\sD$ consists of the even sector of the original theory $\sA$,
coming from both the untwisted and the twisted sector of $\sA$.
We can also assemble the odd sector of the original theory $\sA$, from both the untwisted and twisted sector of $\sA$, into the twisted sector of the theory $\sD$.
This means that the states on $S^1$ of the theory $\sD$
are as shown in Table~\ref{states}.
We easily see that the theory $\sD$ also has a $\bZ_2$ symmetry, and the orbifold of the theory $\sD$ by this $\bZ_2$ regenerates the theory $\sA$ \cite{Vafa:1989ih}.
This $\bZ_2$ gauging is known to be a generalized abstract version of the Kramers-Wannier transformation.

\begin{modified}
The next operation, which is a generalized abstract version of the Jordan-Wigner transformation, uses
the low-energy limit of the nontrivial topological phase of the Kitaev chain \cite{Kitaev:2001kla}. 
This is a fermionic chain whose lowest energy state  on $S^1$ is non-degenerate.
Denoting the fermion parity as $(-1)^F$,
the ground state has $(-1)^F=+1$ when the fermion is antiperiodic around $S^1$,
and has $(-1)^F=-1$ when the fermion is periodic around $S^1$ \cite{Kawabata:2017aa}.
For brevity, we call this topological phase of the Kitaev chain simply ``the Kitaev chain''.

We now consider the theory $\sA \times \Kitaev$ obtained by stacking the Kitaev chain to the original model $\sA$. 
We then take the orbifold by $\bZ_2$,
where the $\bZ_2$ action on the Kitaev chain is given by the fermion parity.
The result is the fermionic model we denote by $\sF$.

To find the decomposition of states of this theory,
we study the four sectors of $\sA\times \Kitaev$,
depending on whether it is untwisted $(u=+1)$ or twisted $(u=-1)$,
and whether the fermion in the final theory $\sF$ is periodic $(s=+1)$ or antiperiodic $(s=-1)$.
It is important to keep in mind that
the actual periodicity of the fermion of the Kitaev chain is given by the product $su$,
since the $\bZ_2$ symmetry we use in the twisting also involves the fermion parity of the Kitaev chain.
This means that $(-1)^F=-su$, due to the property of the Kitaev chain.

Let us denote the $\bZ_2$ charge of $\sA$ by $Q_{\sA}=\pm1$.
Then the total $\bZ_2$ charge is $Q:=Q_{\sA} (-1)^F$.
The theory $\sF$ is obtained by keeping only the states with $Q=+1$.
This means that, to find that the decomposition of states of $\sF$ for each $s=\pm1$,
we simply consider both possibilities $u=\pm1$
and take the states with $Q_{\sA}=-su$, which also equals $(-1)^F$.
The result is summarized in Table~\ref{states}.
There, we refer to states with $(-1)^F=+1$ as bosonic and those with $(-1)^F=-1$ as fermionic.
\end{modified}

We can also perform the same operation against the theory $\sD$, by considering the $\bZ_2$-orbifold of $\sD\times \Kitaev$. 
The decomposition of states of the resulting theory, which we call $\stF$, is also shown in Table~\ref{states}.
In the table, 
we note that $\sF$ and $\stF$ are related simply by exchanging the assignment of $(-1)^F$ in the periodic sector. 
Or equivalently, we have $\stF=\sF\times \Kitaev$.

We can summarize the relation of four theories $\sA$, $\sD$, $\sF$ and $\stF$ in the following diagram: 
\tikzset{>=stealth}
\begin{equation}
\begin{tikzpicture}[baseline=(X.south)]
\node (A) at (0,0) {$\sA$};
\node (D) at (3,0) {$\sD$};
\node (X) at (1.5,-1) {};
\node (F) at (0,-2) {$\vphantom{\stF}\sF$};
\node (tildeF) at (3,-2) {$\stF$};
\draw[<->] (A) to node[above] {\footnotesize $\bZ_2$ orbifold =} 
node[below] {\footnotesize Kramers-Wannier} 
(D);
\draw[<->] (A) to node[left] {\footnotesize Jordan-Wigner} (F);
\draw[<->] (D) to node[right] {\footnotesize Jordan-Wigner}(tildeF);
\draw[<->] (F) to node[above] {\footnotesize $\times\Kitaev$} (tildeF);
\end{tikzpicture}.
\end{equation}

\begin{table*}
\begin{tabular}{c|ccccc}
& \text{A-type models} & & \text{D-type models} & & \text{fermionic models} \\
\hline
$m=3$ & \text{critical Ising} &$=$& \text{critical Ising} & $\leftrightarrow$ & \text{Majorana fermion} \\
$m=4$ & \text{tricritical Ising} &$=$& \text{tricritical Ising} & $\leftrightarrow$ & \text{smallest \Nequals1 minimal model} \\
$m=5$ & \text{tetracritical Ising} &$\leftrightarrow$ & \text{critical 3-state Potts} & $\leftrightarrow$ & \text{fermionic $m=5$ model} \\
\vdots & \vdots & & \vdots &&\vdots
\end{tabular}

\begin{tabular}{ccc}
$m=3$ :&
\begin{tabular}{ll}
\cellcolor{lightblue}$S= 1 + \epsilon\bar\epsilon$,&
\cellcolor{lightpink}$U= \sigma\bar\sigma$, \\
\cellcolor{lightgreen}$T= \sigma\bar\sigma$, &
\cellcolor{lightyellow}$V= \epsilon + \bar\epsilon$,
\end{tabular}&
\begin{tabular}{|c|c|c|c|}
\hline
$s=1$ & $1 (0)$& $\sigma (\frac{1}{16})$& $\epsilon (\frac12)$\\
\hline
& $r=1$ & $r=2$ & $r=3$\\
\hline
\end{tabular}\\
$m=4$ : &
\begin{tabular}{ll}
\cellcolor{lightblue}$S= 1 + \epsilon\bar\epsilon +  \epsilon'\bar\epsilon' +  \epsilon''\bar\epsilon''$,\\
\cellcolor{lightgreen}$T= \sigma\bar\sigma+ \sigma'\bar\sigma'$, \\
\cellcolor{lightpink}$U= \sigma\bar\sigma + \sigma'\bar\sigma'$, \\
\cellcolor{lightyellow}$V= \epsilon'' + \bar\epsilon'' + \epsilon\bar\epsilon'+\epsilon'\bar\epsilon$,
\end{tabular}&
\begin{tabular}{|c|c|c|c|}
\hline
$s=2$ & $\epsilon (\frac1{10})$ & $\sigma (\frac{3}{80})$ & $\epsilon' (\frac35)$\\
\hline
$s=1$ & $1 (0)$ & $\sigma' (\frac7{16})$& $\epsilon'' (\frac32)$\\
\hline
& $r=1$ & $r=2$ & $r=3$\\
\hline
\end{tabular}\\
$m=5$ :&
\begin{tabular}{ll}
\cellcolor{lightblue}$S= 1 + \epsilon\bar\epsilon +  \epsilon'\bar\epsilon' +
 \epsilon''\bar\epsilon'' + \epsilon'''\bar\epsilon'''+\epsilon''''\bar\epsilon''''$,\\
\cellcolor{lightgreen}$T= \sigma\bar\sigma + \sigma'\bar\sigma'+
\sigma''\bar\sigma''+\sigma'''\bar\sigma'''$,\\
\cellcolor{lightpink}$U= \epsilon\bar\epsilon+\epsilon'''\bar\epsilon'+\epsilon'\bar\epsilon'''+
\epsilon''''+\epsilon''\bar\epsilon''+\bar\epsilon''''$, \\
\cellcolor{lightyellow}$V= \sigma\bar\sigma''+\sigma''\bar\sigma+\sigma'\bar\sigma'''+\sigma'''\bar\sigma'$
\end{tabular}&
\begin{tabular}{|c|c|c|c|c|c|}
\hline
$s=2$ & $\epsilon' (\frac25)$& $\sigma (\frac1{40})$ & $\epsilon (\frac1{15})$ & $\sigma'' (\frac{21}{40})$& $\epsilon''' (\frac75)$ \\
\hline
$s=1$ & $1 (0)$& $\sigma' (\frac18)$ & $\epsilon'' (\frac23)$  & $\sigma''' (\frac{13}8)$ & $\epsilon'''' (3)$\\
\hline
& $r=1$ & $r=2$ & $r=3$ & $r=4$ & $r=5$\\
\hline
\end{tabular}
\end{tabular}
\caption{Details of the $m=3,4,5$ models.
\begin{modified}
The upper part gives conventional names for the models when available.
The lower left part lists the Virasoro content of the states $S$, $T$, $U$ and $V$.
The lower right part provides the mapping between the symbols $\epsilon$, $\sigma$ etc.~and the characters $\chi_{r,s}$.
For example, $\epsilon(\frac1{10})$ for $m=2$, $r=1$, $s=2$ means that 
we use the entry $\epsilon$ for $\chi_{1,2}$ whose primary has $L_0=\frac1{10}$.
\end{modified}
\label{examples}}
\end{table*}

\subsection{Application to the unitary minimal models}
Let us now recall the well-known fact that the D-type modular invariants are obtained by a $\bZ_2$ orbifold, or equivalently a $\bZ_2$-gauging, of the A-type modular invariants. This means that we can apply the general method explained above to produce fermionic  minimal models. 
Explicitly, these models have the following operator content.

We denote the irreducible Virasoro characters at $c=1-\frac{6}{m(m+1)}$ by $\chi_{r,s}$.
We set $p=m+1$ and $q=m$  when $m$ is even,
and $q=m+1$ and $p=m$ when $m$ is odd.
This is to make  $p$  always odd and $q$   always even.
We then let $1\le r\le q-1$ and $1\le s\le p-1$.
The conformal weight $L_0$ of $\chi_{r,s}$ is then given by $
L_0 = \frac{(pr-qs)^2-1}{4pq}.
$
This set is redundant because of the two-fold identification $\chi_{r,s}=\chi_{q-r,p-s}$.
We remove this redundancy by restricting  $s\le (p-1)/2$.
We then have 
\begin{equation}
\hskip-1em\begin{array}{ll}
\displaystyle\cellcolor{lightblue}S= \sum_{r\equiv 1}\sum_{s}  \chi_{r,s} \overline{\chi_{r,s}},&
\displaystyle\cellcolor{lightpink}U= \sum_{r\equiv \frac{q}2}\sum_{s}  \chi_{r,s} \overline{\chi_{q-r,s}},\\
\displaystyle\cellcolor{lightgreen}T= \sum_{r\equiv 0}\sum_{s}  \chi_{r,s} \overline{\chi_{r,s}},&
\displaystyle\cellcolor{lightyellow}V= \sum_{r\equiv \frac{q}2+1}\sum_{s}  \chi_{r,s} \overline{\chi_{q-r,s}},
\end{array}
\label{STUV}
\end{equation}
where we abused the notation and identified a state space and its character;
$a\equiv b$ is the equality modulo 2.
We note that the parity of $r\equiv q/2$ for $U$ and $r\equiv q/2+1$ for $V$ is correlated to the spin of the states being integral or half-integral.

\begin{modified}
The expressions \eqref{STUV}  can be obtained as follows.
By definition, $S+T$ and $S+U$ are equal to the partition functions of the A-type and the D-type minimal model, which can be found in the standard textbooks on 2d conformal field theory, e.g.~\cite{Francesco:2012aa}.
By performing a modular transformation $\tau\to-1/\tau$, one then obtains $U+V$ and $T+V$, respectively.
From this information we can extract $S$, $T$, $U$ and $V$ individually.
These expressions can also be obtained from  a very general result of Ref. \cite{Petkova:2000ip} as applied to the minimal models.
\end{modified}

The spectra for  $m=3,4,5$ are shown in Table~\ref{examples}.
We used $\epsilon$ for $\bZ_2$-even primaries and $\sigma$ for $\bZ_2$-odd primaries;
those with larger $L_0$ have more primes in the superscript.
The operators in $S$, $T$, and $U$  all have integer spins,
while the operators in $V$ all have half-integral spins.
For $m=3,4$, we have $U=T$, meaning that there is no distinction between A-type and D-type models.
For $m=5$, $U\neq T$, and the A-type model and the D-type model are distinct.

For $m=3$, $\epsilon$ is the free fermion with spin $1/2$;
for $m=4$, $\epsilon''$ is the supersymmetry generator in the fermionic model;
for $m=5$, $\epsilon''''$ is the $W_3$ generator and exists in the untwisted sector of the D-type model.
The pattern repeats itself. 
We find that the chiral algebra of the D-type model for $m\equiv 5,6$ mod 4 has a W-generator of integer spin,
and that the chiral algebra of the fermionic model for $m\equiv 3,4$ mod 4 has a W-generator of half-integral spin, as was mentioned in Ref. \cite{Runkel:2020zgg}.

\section{Analysis with the spin chain}\label{sec:chain}

\subsection{General analysis}
Let us begin by recalling the  Jordan-Wigner transformation of a spin-1/2 chain 
\footnote{%
The Kramers-Wannier transformation of a spin-1/2 chain is discussed in the Supplemental Material.}.
We consider a circular chain with sites labeled by a positive integer $i$, each hosting the local Hilbert space $\bC^2$.
We denote the local Pauli matrices as $\sigma_{x,y,z}^{(i)}$,
and consider the on-site $\bZ_2$ symmetry generated by $\sigma_z$, so that the global $\bZ_2$ charge is given by $\prod \sigma_z^{(i)}$.

The Jordan-Wigner transformation is given by the following relation \begin{equation}
\begin{aligned}
\psi^{(2i-1)} &:= (\prod_{1\le j< i} \sigma_z^{(j)}) \sigma_x^{(i)},\\
\psi^{(2i)} &:= (\prod_{1\le j< i} \sigma_z^{(j)}) \sigma_y^{(i)}.
\end{aligned}
\end{equation}
This is a non-local transformation, but maps local operators to local operators when restricted to $\bZ_2$-even and/or bosonic operators.
To see this, we note that any $\bZ_2$-even operator can be generated from $\sigma_z^{(i)}$ and $\sigma_x^{(i)}\sigma_x^{(i+1)}$, and that they are mapped by the Jordan-Wigner transformation as follows: \begin{equation}
\begin{aligned}
\sigma_z^{(i)} &=  -\ii\psi^{(2i-1)}\psi^{(2i)},\\
\sigma_x^{(i)}\sigma_x^{(i+1)} &=-\ii \psi^{(2i)}\psi^{(2i+1)}.
\end{aligned}\label{foo}
\end{equation}

Let us now show that this mapping reproduces the general analysis in continuum theory
when we consider a circular chain of $N$ sites.
\begin{modified}
If we impose the boundary condition $
\psi^{(2N+1)} = s \psi^{(1)}\label{f}
$ 
where $s=\pm1$,
then the relation \eqref{foo} is slightly modified when $i=N$ to be \begin{equation}
-\ii\psi^{(2N)} \psi^{(2N+1)} = -s (\prod_i \sigma_z^{(i)}) \sigma_x^{(N)}\sigma_x^{(1)}.
\end{equation}
The right-hand side should equal $\sigma_x^{(N)} \sigma_x^{(N+1)}$.
This means that the periodicity of the original spin chain
is given by $\sigma_x^{(N+1)}=( -st ) \sigma_x^{(1)}$,
where $s$ is the sign determining the periodicity of the fermion chain as above
and $t$ is the global $\bZ_2$ charge $\prod_i \sigma_z^{(i)}$.
It is also clear that the global $\bZ_2$ charge agrees with the fermion parity $\prod_i (-\ii\psi^{(2i-1)}\psi^{(2i)})$.
\end{modified}
This explains the mapping of states between the original theory $\sA$ 
and the fermionized theory $\sF$. 

We now note that the relation between the theory $\sF$ and the theory $\stF$ can be realized at the level of the fermion chain by the shift $\psi^{(i)} \to \psi^{(i+1)}$. 
Indeed, when the boundary condition is given by $\psi^{(2N+1)} = s \psi^{(1)}$ where $s=\pm1$,
 the fermion number operator after the shift is 
 \begin{equation}
\psi^{(2)} \cdots \psi^{(2N)} \psi^{(2N+1)} 
= -s \psi^{(1)} \psi^{(2)} \cdots \psi^{(2N)}. 
\end{equation}
This means that the fermion number assignment gets reversed only in the periodic sector.

\subsection{Application to the unitary minimal models}

To obtain a Majorana chain realizing the fermionic minimal models,
we simply need to take a realization of ordinary bosonic minimal models on the spin-1/2 chain with the manifest $\bZ_2$ symmetry, and perform the Jordan-Wigner transformation.

This method is well known to work for the Ising model and the tricritical Ising model.
There are two apparent obstacles to generalize this construction to higher minimal models:
\begin{modified}
i) Some of these known bosonic models do not have manifest $\bZ_2$ symmetry, while the $\bZ_2$ symmetry emerges only in the long-range limit (see, e.g., \cite{DCruz:2005aa}).
ii) Most of the known bosonic models realizing the ordinary minimal models higher than these are defined on a chain of ``spins'' larger than $1/2$.
That is, they are realized on a generalized spin chain such that each site has the state space $\bC^k$ with $k>2$.
While we currently do
not have any solutions to the first point, the second point
can be easily circumvented.

\end{modified}

Suppose we are given a spin-chain Hamiltonian realizing a higher minimal model
with an explicit $\bZ_2$ symmetry 
\footnote{
\begin{modified}
Luckily, such models are available for all $m$.
Indeed, we take the restricted solid-on-solid (RSOS) model \cite{Andrews:1984af} and its generalization \cite{Pasquier:1986jc}, which are known to reproduce minimal models at criticality \cite{Huse:1984mn}.
Such a model is specified by a Dynkin diagram $\Gamma$, such that
the states at each site of the corresponding ``spin'' chain are labeled by the nodes of $\Gamma$
and that the Hamiltonian are specified by the edges of $\Gamma$.
Then the symmetry of the Dynkin diagram $\Gamma$ becomes the symmetry of the Hamiltonian.
The A-type minimal model is a special case of this construction where $\Gamma$ is the $A_m$ Dynkin diagram, and our $\bZ_2$ symmetry is the symmetry flipping this diagram.
In passing we note that exactly the same  RSOS spin chains  for the A-type minimal model 
can also be realized in terms of interacting chains of anyons \cite{Feiguin:2006ydp}.
\end{modified}
}
such that the state space at each site is $\bC^k$.
We pick an integer $\ell$ so that we can embed $\bC^k \subset (\bC^2)^{\otimes \ell}$,
i.e.~we represent one site of the original spin chain in terms of a unit cell consisting of $\ell$ sites of the spin-1/2 chain.
It is clear that this can be done in a way preserving the $\bZ_2$ symmetry.
When $k$ is not a power of two, we have $2^\ell-k$ unnecessary states after the embedding,
but they can be removed by adding to the Hamiltonian a local term which gives a very large energy to these unnecessary states.
Then the low-lying states before and after the embedding into the spin-1/2 chain are effectively the same, and eventually we will have a local Hamiltonian on the spin-1/2 chain with a manifest $\bZ_2$ symmetry realizing the higher minimal model.

Let us illustrate this procedure by taking the 3-state Potts model. The standard Hamiltonian realization of the 3-state Potts model is on a spin chain with each site having $\bC^3$,
with the basis $\ket{A}$, $\ket{B}$, and $\ket{C}$,
acted on by the clock and shift operators 
\begin{equation}
\begin{aligned}
Z: \ket{A}&\to \ket{A}, \ \ket{B}\to \omega\ket{B},\  \ket{C} \to \bar\omega \ket{C},\\
X: \ket{A} &\to \ket{B} \to \ket{C} \to \ket{A}
\end{aligned}
\end{equation}
where $\omega=e^{2\pi i/3}$.
The Hamiltonian is then  \begin{equation}
H_{\text{Potts}}= - \sum_i  ( J Z_{(i)} Z_{(i+1)}^{-1} + f X_{(i)} + h.c. )
\end{equation}
where $J$ and $f$ are parameters,
and the $\bZ_2$ symmetry is generated by $ 
\ket{A} \to \ket{A}, \ 
\ket{B} \leftrightarrow \ket{C}
$ 
at every site. 
The model becomes critical when $J=f$.

We now embed $\bC^3$ to $(\bC^2)^{\otimes2}$ by choosing 
\begin{equation}
\begin{aligned}
\ket{A}_i&=\ket{\uparrow}_{2i-1}\ket{\uparrow}_{2i},\\
\ket{B}_i&=\frac1{\sqrt{2}}\ket{\downarrow}_{2i-1}(+\ket{\uparrow}_{2i} + \ket{\downarrow}_{2i}), \\
\ket{C}_i&=\frac1{\sqrt{2}}\ket{\downarrow}_{2i-1}(-\ket{\uparrow}_{2i} + \ket{\downarrow}_{2i}).
\end{aligned}
\label{po}
\end{equation}
This preserves the $\bZ_2$ symmetry. 
We also note that we have one unnecessary state $\ket{D}_{i}:=\ket{\uparrow}_{2i-1}\ket{\downarrow}_{2i}$. 
This state can be removed by adding to the Hamiltonian \begin{equation}
\ket{D}_i\bra{D}_i= \frac14(1+\sigma_z^{(2i-1)})(1-\sigma_z^{(2i)})\label{U}
\end{equation} with a huge positive coefficient $U$.

Using \eqref{po} and \eqref{U}, we can rewrite $H_{\text{Potts}}$ in terms of Pauli matrices.
\begin{modified}
The 3-state Potts model is thus translated to a model on a spin-1/2 chain
constructed from $\sigma_z$ and bilinears of $\sigma_{x,y}$,
which can be Jordan-Wigner transformed into a chain of interacting Majorana fermions, with the following Hamiltonian
\begin{widetext}
\begin{align}
H &= JH_1 + fH_2 + UH_3, \nonumber\\
H_1 &=
-\frac{1}{8}\sum_i  \Bigl[ 
(2\ii\psi^{(4i-3)}\psi^{(4i-2)}+\ii\psi^{(4i-1)}\psi^{(4i)}+\psi^{(4i-3)}\psi^{(4i-2)}\psi^{(4i-1)}\psi^{(4i)})
 \nonumber \\
&\quad\quad\quad\quad\quad\cdot(2\ii\psi^{(4i+1)}\psi^{(4i+2)}+\ii\psi^{(4i+3)}\psi^{(4i+4)}+\psi^{(4i+1)}\psi^{(4i+2)}\psi^{(4i+3)}\psi^{(4i+4)})  \nonumber \\
&\quad\quad\quad\quad\quad + 3(1+\ii\psi^{(4i-3)}\psi^{(4i-2)})\ii\psi^{(4i)}\psi^{(4i+1)}(1-\ii\psi^{(4i+1)}\psi^{(4i+2)})\ii\psi^{(4i+2)}\psi^{(4i+3)}  
\Bigr], \nonumber\\
H_2 &= -\frac{1}{2}\sum_i  (1+\ii\psi^{(4i-3)}\psi^{(4i-2)})\ii\psi^{(4i-1)}\psi^{(4i)} +\frac{1}{\sqrt2}\sum_i   (\ii\psi^{(4i-2)}\psi^{(4i-1)}-\ii\psi^{(4i-3)}\psi^{(4i)}), \nonumber\\
H_3 &= -\frac{1}{4}\sum_i  (1-\ii\psi^{(4i-3)}\psi^{(4i-2)})(1+\ii\psi^{(4i-1)}\psi^{(4i)}). \label{m=5_fMM}
\end{align}
\end{widetext}
As the critical 3-state Potts model gives the $m=5$ D-type modular invariant, the Majorana chain \eqref{m=5_fMM} at criticality will give the $m=5$ fermionic minimal model
\footnote{
One can also do the Kramers-Wannier transformation on the spin-1/2 chain converted from the 3-state Potts model, which will end up with another spin-1/2 chain whose critical point corresponds to the $m=5$ A-type modular invariant, i.e.~the tetracritical Ising model.
}.
We numerically checked that the Hamiltonian \eqref{m=5_fMM} does give a conformal field theory with $c=4/5$ when $J=f$
\footnote{
For details, see the Supplemental Material.
}.
It would also be interesting to study in detail e.g.~the two-point functions of the fermionic operators, from which we could identify their scaling dimensions which have been indicated in TABLE \ref{examples} for the $m=5$ fermionic minimal model. 
We leave it to a future work.
\end{modified}

\paragraph{Acknowledgments:} 
The authors thank Hosho Katsura and Ryohei Kobayashi for their comments on this Letter.
CTH and YT are in part supported  by WPI Initiative, MEXT, Japan at IPMU, the University of Tokyo.
CTH is also supported in part by JSPS KAKENHI Grant No.19K14608.
YN is supported in part by JSPS KAKENHI Grant No.17K14301.
YT is also supported in part by JSPS KAKENHI Grant No.17H04837 
and No.16H06335.

\bibliographystyle{ytphys}
\baselineskip=.95\baselineskip
\bibliography{ref}

\providecommand{\href}[2]{#2}\begingroup\raggedright\begin{thebibliography}{10}

\bibitem{Belavin:1984aa}
A.~A. Belavin, A.~M. Polyakov, and A.~B. Zamolodchikov, {\slshape Infinite
  conformal symmetry in two-dimensional quantum field theory,}
  \href{http://www.sciencedirect.com/science/article/pii/055032138490052X}{{\em
  Nuclear Physics B} {\bfseries 241} (1984) 333--380}.

\bibitem{Friedan:1984aa}
D.~Friedan, Z.~Qiu, and S.~Shenker, {\slshape {Conformal Invariance, Unitarity,
  and Critical Exponents in Two Dimensions},}
  \href{https://link.aps.org/doi/10.1103/PhysRevLett.52.1575}{{\em Phys. Rev.
  Lett.} {\bfseries 52} (1984) 1575}.

\bibitem{Cardy:1986aa}
J.~L. Cardy, {\slshape Operator content of two-dimensional conformally
  invariant theories,}
  \href{http://www.sciencedirect.com/science/article/pii/0550321386905523}{{\em
  Nuclear Physics B} {\bfseries 270} (1986) 186--204}.

\bibitem{Kato:1987td}
A.~Kato, {\slshape {Classification of Modular Invariant Partition Functions in
  Two-Dimensions},}
\href{http://dx.doi.org/10.1142/S0217732387000732}{{\em Mod. Phys. Lett.}
  {\bfseries A02} (1987) 585}.

\bibitem{Cappelli:1986hf}
A.~Cappelli, C.~Itzykson, and J.~B. Zuber, {\slshape {Modular Invariant
  Partition Functions in Two-Dimensions},}
\href{http://dx.doi.org/10.1016/0550-3213(87)90155-6}{{\em Nucl. Phys.}
  {\bfseries B280} (1987) 445--465}.

\bibitem{Cappelli:1987xt}
A.~Cappelli, C.~Itzykson, and J.~B. Zuber, {\slshape {The ADE Classification of
  Minimal and {$A^1_{(1)}$} Conformal Invariant Theories},}
\href{http://dx.doi.org/10.1007/BF01221394}{{\em Commun. Math. Phys.}
  {\bfseries 113} (1987) 1}.

\bibitem{Andrews:1984af}
G.~Andrews, R.~Baxter, and P.~Forrester, {\slshape {Eight Vertex Sos Model and
  Generalized Rogers-Ramanujan Type Identities},}
  \href{http://dx.doi.org/10.1007/BF01014383}{{\em J. Statist. Phys.}
  {\bfseries 35} (1984) 193--266}.

\bibitem{Huse:1984mn}
D.~A. Huse, {\slshape {Exact Exponents for Infinitely Many New Multicritical
  Points},} \href{http://dx.doi.org/10.1103/PhysRevB.30.3908}{{\em Phys. Rev.
  B} {\bfseries 30} (1984) 3908--3915}.

\bibitem{Pasquier:1986jc}
V.~Pasquier, {\slshape {Two-Dimensional Critical Systems Labelled by Dynkin
  Diagrams},} \href{http://dx.doi.org/10.1016/0550-3213(87)90332-4}{{\em Nucl.
  Phys. B} {\bfseries 285} (1987) 162--172}.

\bibitem{Fendley:2006aa}
P.~Fendley, {\slshape Loop models and their critical points,}
  \href{http://dx.doi.org/10.1088/0305-4470/39/50/011}{{\em J. Phys. A Math.
  Gen.} {\bfseries 39} (2006) 15445--15475}.

\bibitem{Feiguin:2006ydp}
A.~Feiguin, S.~Trebst, A.~W.~W. Ludwig, M.~Troyer, A.~Kitaev, Z.~Wang, and
  M.~H. Freedman, {\slshape {Interacting Anyons in Topological Quantum Liquids:
  the Golden Chain},}
  \href{http://dx.doi.org/10.1103/PhysRevLett.98.160409}{{\em Phys. Rev. Lett.}
  {\bfseries 98} (2007) 160409}, \href{http://arxiv.org/abs/cond-mat/0612341}{{
  arXiv:cond-mat/0612341}}.

\bibitem{SCHULTZ:1964aa}
T.~D. Schultz, D.~C. Mattis, and E.~H. Lieb, {\slshape {Two-Dimensional Ising
  Model as a Soluble Problem of Many Fermions},}
  \href{https://link.aps.org/doi/10.1103/RevModPhys.36.856}{{\em Rev. Mod.
  Phys.} {\bfseries 36} (1964) 856}.

\bibitem{Friedan:1985aa}
D.~Friedan, Z.~Qiu, and S.~Shenker, {\slshape {Superconformal invariance in two
  dimensions and the tricritical Ising model},}
  \href{http://www.sciencedirect.com/science/article/pii/0370269385908196}{{\em
  Phys. Lett. B} {\bfseries 151} (1985) 37}.

\bibitem{Rahmani:2015aa}
A.~Rahmani, X.~Zhu, M.~Franz, and I.~Affleck, {\slshape {Emergent Supersymmetry
  from Strongly Interacting Majorana Zero Modes},}
  \href{https://link.aps.org/doi/10.1103/PhysRevLett.115.166401}{{\em Phys.
  Rev. Lett.} {\bfseries 115} (2015) 166401}.

\bibitem{Note1}
The authors learned very recently that such theories were found in the late
  1980s \cite {Petkova:1988cy,Furlan:1989ra}. They thank Prof.~V. Petkova for
  information.

\bibitem{YTCernLect}
Y.~Tachikawa, {\slshape Topological phases and relativistic {QFTs},}
  \url{https://member.ipmu.jp/yuji.tachikawa/lectures/2018-cern-rikkyo/}. Notes
  of the lectures given in the CERN winter school, February 2018.

\bibitem{Karch:2019lnn}
A.~Karch, D.~Tong, and C.~Turner, {\slshape {A Web of 2d Dualities: ${\bf Z}_2$
  Gauge Fields and Arf Invariants},}
  \href{http://dx.doi.org/10.21468/SciPostPhys.7.1.007}{{\em SciPost Phys.}
  {\bfseries 7} (2019) 007},
\href{http://arxiv.org/abs/1902.05550}{{ arXiv:1902.05550~[hep-th]}}.

\bibitem{Gaiotto:2015zta}
D.~Gaiotto and A.~Kapustin, {\slshape {Spin TQFTs and Fermionic Phases of
  Matter},} \href{http://dx.doi.org/10.1142/S0217751X16450445}{{\em Int. J.
  Mod. Phys.} {\bfseries A31} (2016) 1645044},
\href{http://arxiv.org/abs/1505.05856}{{ arXiv:1505.05856~[cond-mat.str-el]}}.

\bibitem{Bhardwaj:2016clt}
L.~Bhardwaj, D.~Gaiotto, and A.~Kapustin, {\slshape {State Sum Constructions of
  Spin-TFTs and String Net Constructions of Fermionic Phases of Matter},}
\href{http://arxiv.org/abs/1605.01640}{{ arXiv:1605.01640~[cond-mat.str-el]}}.

\bibitem{Vafa:1989ih}
C.~Vafa, {\slshape {Quantum Symmetries of String Vacua},}
\href{http://dx.doi.org/10.1142/S0217732389001842}{{\em Mod. Phys. Lett.}
  {\bfseries A04} (1989) 1615}.

\bibitem{Kitaev:2001kla}
A.~Y. Kitaev, {\slshape {Unpaired Majorana fermions in quantum wires},}
  \href{http://dx.doi.org/10.1070/1063-7869/44/10S/S29}{{\em Phys. Usp.}
  {\bfseries 44} (2001) 131--136},
\href{http://arxiv.org/abs/cond-mat/0010440}{{
  arXiv:cond-mat/0010440~[cond-mat.mes-hall]}}.

\bibitem{Kawabata:2017aa}
K.~Kawabata, R.~Kobayashi, N.~Wu, and H.~Katsura, {\slshape Exact zero modes in
  twisted kitaev chains,}
  \href{https://ui.adsabs.harvard.edu/abs/2017PhRvB..95s5140K}{{\em Phys. Rev.
  B} {\bfseries 95} (2017) 195140}, \href{http://arxiv.org/abs/1702.00197}{{
  arXiv:1702.00197~[cond-mat.mes-hall]}}.

\bibitem{Francesco:2012aa}
P.~Francesco, P.~Mathieu, and D.~S{\'e}n{\'e}chal, {\em Conformal field
  theory}.
\newblock Springer Science \& Business Media, 2012.

\bibitem{Petkova:2000ip}
V.~B. Petkova and J.~B. Zuber, {\slshape {Generalized Twisted Partition
  Functions},} \href{http://dx.doi.org/10.1016/S0370-2693(01)00276-3}{{\em
  Phys. Lett.} {\bfseries B504} (2001) 157--164},
\href{http://arxiv.org/abs/hep-th/0011021}{{ arXiv:hep-th/0011021}}.

\bibitem{Runkel:2020zgg}
I.~Runkel and G.~M.~T. Watts, {\slshape {Fermionic CFTs and Classifying
  Algebras},}
\href{http://arxiv.org/abs/2001.05055}{{ arXiv:2001.05055~[hep-th]}}.

\bibitem{Note2}
The Kramers-Wannier transformation of a spin-1/2 chain is discussed in the
  Supplemental Material.

\bibitem{DCruz:2005aa}
C.~D'Cruz and J.~K. Pachos, {\slshape {Chiral phase from three-spin
  interactions in an optical lattice},}
  \href{https://link.aps.org/doi/10.1103/PhysRevA.72.043608}{{\em Phys. Rev. A}
  {\bfseries 72} (2005) 043608}.

\bibitem{Note3}
\begin {modified} Luckily, such models are available for all $m$. Indeed, we
  take the restricted solid-on-solid (RSOS) model \cite {Andrews:1984af} and
  its generalization \cite {Pasquier:1986jc}, which are known to reproduce
  minimal models at criticality \cite {Huse:1984mn}. Such a model is specified
  by a Dynkin diagram $\Gamma $, such that the states at each site of the
  corresponding ``spin'' chain are labeled by the nodes of $\Gamma $ and that
  the Hamiltonian are specified by the edges of $\Gamma $. Then the symmetry of
  the Dynkin diagram $\Gamma $ becomes the symmetry of the Hamiltonian. The
  A-type minimal model is a special case of this construction where $\Gamma $
  is the $A_m$ Dynkin diagram, and our $\protect \mathbb {Z}_2$ symmetry is the
  symmetry flipping this diagram. In passing we note that exactly the same RSOS
  spin chains for the A-type minimal model can also be realized in terms of
  interacting chains of anyons \cite {Feiguin:2006ydp}. \end {modified}.

\bibitem{Note4}
One can also do the Kramers-Wannier transformation on the spin-1/2 chain
  converted from the 3-state Potts model, which will end up with another
  spin-1/2 chain whose critical point corresponds to the $m=5$ A-type modular
  invariant, i.e.~the tetracritical Ising model.

\bibitem{Note5}
For details, see the Supplemental Material.

\bibitem{Petkova:1988cy}
V.~B. Petkova, {\slshape {Two-dimensional (Half) Integer Spin Conformal
  Theories With Central Charge $c < 1$},}
  \href{http://dx.doi.org/10.1142/S0217751X88001235}{{\em Int. J. Mod. Phys. A}
  {\bfseries 3} (1988) 2945--2958}.

\bibitem{Furlan:1989ra}
P.~Furlan, A.~C. Ganchev, and V.~B. Petkova, {\slshape {Fusion Matrices and $c
  < 1$ (Quasi)local Conformal Theories},}
  \href{http://dx.doi.org/10.1142/S0217751X90001252}{{\em Int. J. Mod. Phys. A}
  {\bfseries 5} (1990) 2721--2736}. [Erratum: Int.J.Mod.Phys.A 5, 3641 (1990)].

\bibitem{ITensor}
{\slshape {ITensor Library version 3},} \url{http://itensor.org}.

\bibitem{Holzhey:1994we}
C.~Holzhey, F.~Larsen, and F.~Wilczek, {\slshape {Geometric and Renormalized
  Entropy in Conformal Field Theory},}
  \href{http://dx.doi.org/10.1016/0550-3213(94)90402-2}{{\em Nucl. Phys.}
  {\bfseries B424} (1994) 443--467},
\href{http://arxiv.org/abs/hep-th/9403108}{{ arXiv:hep-th/9403108}}.

\bibitem{Calabrese:2004eu}
P.~Calabrese and J.~L. Cardy, {\slshape {Entanglement Entropy and Quantum Field
  Theory},} \href{http://dx.doi.org/10.1088/1742-5468/2004/06/P06002}{{\em J.
  Stat. Mech.} {\bfseries 0406} (2004) P06002},
\href{http://arxiv.org/abs/hep-th/0405152}{{ arXiv:hep-th/0405152}}.

\end{thebibliography}\endgroup

\newpage

\onecolumngrid
\appendix
\section{Supplemental Material}

\subsection{Kramers-Wannier transformation on a circular chain}
Here, we discuss the Kramers-Wannier transformation on a circular chain.
One point to be aware of is that the often-found expressions \begin{equation}
\tilde \sigma_z^{(i)}:=\sigma_x^{(i)}\sigma_x^{(i+1)},\qquad
\tilde \sigma_x^{(i)}:=\prod_{i<j}\sigma_z^{(j)} \label{KW}
\end{equation} do not reproduce the $\bZ_2$-odd states, 
since the expressions above imply $\prod \tilde\sigma_z^{(i)}=1$. 
A consistent transformation is given as follows: \begin{align}
\tilde\sigma_z^{(i)}&:=\sigma_x^{(i)} \sigma_x^{(i+1)},&
\tilde \sigma_z^{(N)} &:=u\sigma_x^{(N)} \sigma_x^{(1)} (\prod_j \sigma_z^{(j)}),\\
\tilde \sigma_x^{(i)} &:=\modi{\ii\,}\sigma_x^{(1)} \prod_{j\le i} \sigma_z^{(j)}, &
\tilde \sigma_x^{(N)} &:=\modi{\ii\,}\sigma_x^{(1)} (\prod_{j} \sigma_z^{(j)})
\end{align}
where $i<N$  and $u=\pm1$ is a sign we will choose later.
We can then deduce \begin{align}
\tilde\sigma_x^{(i-1)}\tilde\sigma_x^{(i)}&=\sigma_z^{(i)},&
\tilde\sigma_x^{(N)}\tilde\sigma_x^{(1)} &= \sigma_z^{(1)} \prod_{j} \sigma_z^{(j)} 
\end{align} where $i\ge 2$ for the first equation. We also have \begin{equation}
\prod \tilde \sigma_z^{(i)} = u \prod \sigma_z^{(i)}.\label{u}
\end{equation}

In order for the Hamiltonian eigenvalues to be unchanged under the Kramers-Wannier transformation, we would like to keep the mapping \begin{align}
\tilde \sigma_z^{(N)} &=\sigma_x^{(N)} \sigma_x^{(N+1)} , &
\tilde\sigma_x^{(N)}\tilde\sigma_x^{(N+1)} &= \sigma_z^{(1)}.\label{keep}
\end{align}
Expressing the boundary conditions of the original chain and the dual chain by two signs $s,t=\pm1$ as \begin{equation}
\sigma_x^{(N+1)}:=s\sigma_x^{(1)},\qquad
\tilde\sigma_x^{(N+1)}:=t\tilde\sigma_x^{(1)},
\end{equation}
we find that the required relation \eqref{keep} can be achieved if we restrict the Hilbert spaces to the sectors satisfying \begin{equation}
u=s\prod \sigma_z^{(i)},\qquad
1=t\prod \sigma_z^{(i)}.
\end{equation} Combining with \eqref{u}, we find \begin{equation}
s=\prod \tilde\sigma_z^{(i)},\qquad
t=\prod \sigma_z^{(i)},
\end{equation} reproducing the mapping between the theory $\sA$ and the theory $\sD$ discussed in the main text.

\subsection{Numerical check of $c$ of the critical point of \eqref{m=5_fMM}}
\begin{modified}
Here we outline our numerical check that the interacting Majorana chain \eqref{m=5_fMM} at the critical point $J=f$ has $c=4/5$. Recall that the Hamiltonian \eqref{m=5_fMM} before the Jordan-Wigner transformation is a chain of spin-$1/2$ sites with the following Hamiltonian (the converted 3-state Potts model):
\end{modified}
\begin{multline}
H= \sum_i \Bigl( \frac U4(1+\sigma_z^{(2i-1)})(1-\sigma_z^{(2i)}) 
- f (\frac12(\sigma_z^{(2i-1)}-1) \sigma_z^{(2i)} + \frac{1}{\sqrt2} (\sigma_x^{(2i-1)}\sigma_x^{(2i)}-\sigma_y^{(2i-1)}\sigma_y^{(2i)}) )\\
-2J \Bigl[ 
(\frac12 \sigma_z^{(2i-1)} +\frac14 \sigma_z^{(2i)} +
\frac14 \sigma_z^{(2i-1)}\sigma_z^{(2i)})
(\frac12 \sigma_z^{(2i+1)} +\frac14 \sigma_z^{(2i+2)} +
\frac14 \sigma_z^{(2i+1)}\sigma_z^{(2i+2)})   \\
 + \frac{\sqrt{3}}{4} (  1- \sigma_z^{(2i-1)} )\sigma_x^{(2i)}    
\frac{\sqrt{3}}{4} (  1- \sigma_z^{(2i+1)} )\sigma_x^{(2i+2)}    
\Bigr]\Bigr).
\label{convertedH}
\end{multline}

\begin{figure}
\centering
\includegraphics[width=.6\textwidth]{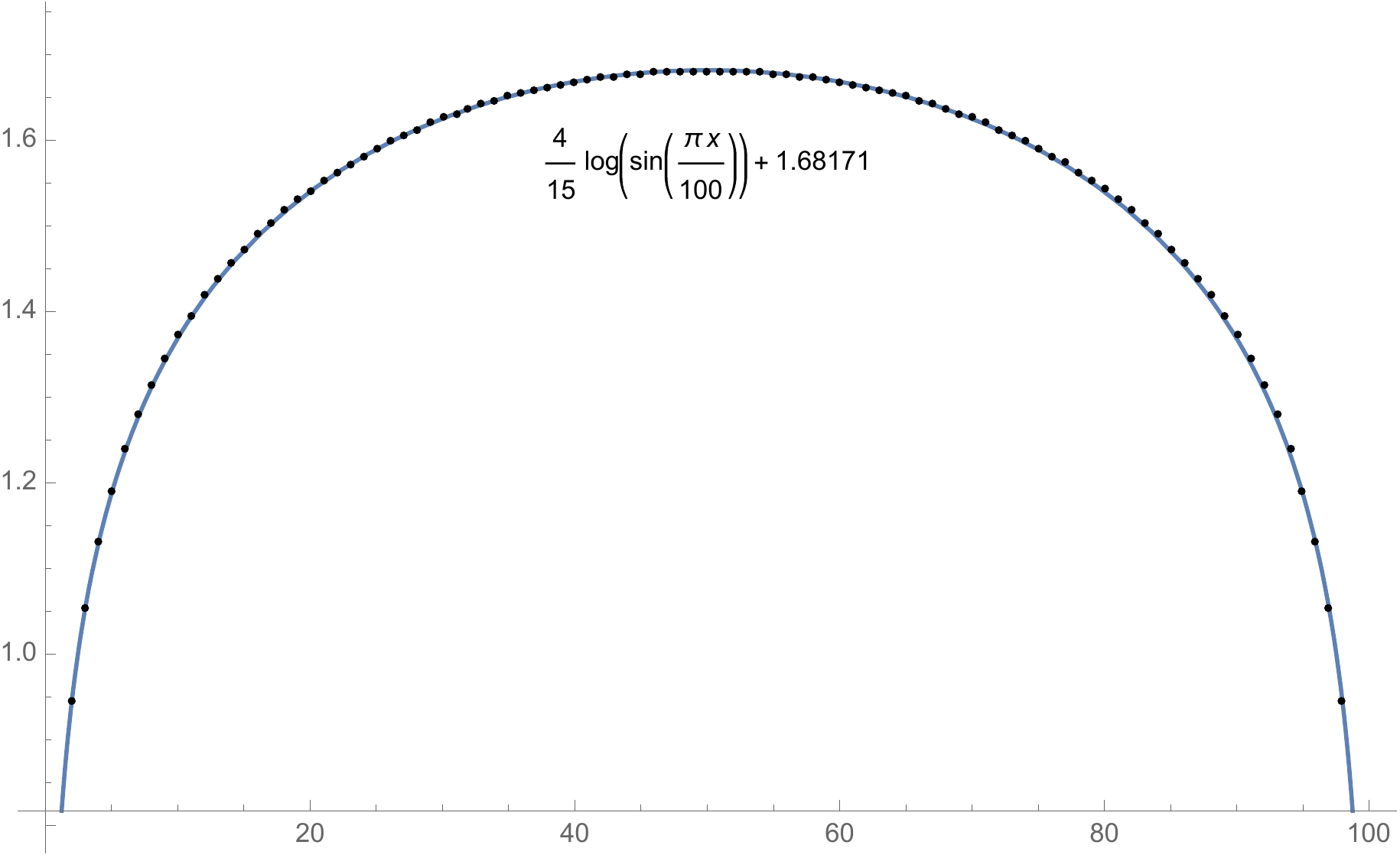}
\caption{The entanglement entropy of the converted 3-state Potts model at $J=f=1$, $U=10^4$, on a chain of $N=100$ pairs of spin $1/2$ sites.
The dots show the numerically computed values; the curved line is the theoretical prediction.
\label{EE}}
\end{figure}

We used the \texttt{ITensor} library \cite{ITensor} to find numerically the ground state wavefunction
on a periodic chain of $N=100$ pairs of spin-$1/2$ sites.
We chose $J=f=1$ to have the critical chain, and chose the coefficient $U$ to project out the unnecessary state $\ket{D}\bra{D}$ to be $U=10^4$.
We then computed the entanglement entropy of the sites $1,2,\ldots, 2i-1,2i$ for $i=1,\ldots,N$ with respect to the complement.

\begin{modified}
We note that our \texttt{ITensor} code is written in terms of the Hamiltonian \eqref{convertedH} using the spin variables, rather than the fermionized version \eqref{m=5_fMM}.
Still, thanks to our general analysis given in the main text, the spectrum of the untwisted even sector of \eqref{convertedH} is completely the same as the spectrum of the antiperiodic bosonic sector of \eqref{m=5_fMM}, which contains the vacuum.
Therefore, our numerical computation does study the property of the fermionic chain.
\end{modified}

The result is plotted in Fig.~\ref{EE}, against the theoretical result \cite{Holzhey:1994we,Calabrese:2004eu}
\begin{equation}
\frac{c}3\log( \sin \frac{\pi x}{N})  + d
\end{equation}
where $d$ is a UV regularization-dependent constant. 
We see a good agreement with the theoretical value $c=4/5$,
where the constant part was fitted numerically.

Before presenting the \texttt{ITensor} code, we would like to point out a few caveats in the numerical computation:
\begin{itemize}
\item The coefficient $U$ in front of $\ket{D}\bra{D}$ can theoretically be taken to be $U\to +\infty$. 
Numerically this needs to be avoided, because this leads to heavy loss of significance in the floating-point computation.
A reasonable choice is as follows.
Assuming $J$ and $f$ to be order 1, the energy due to the $J$ and $f$ terms is of order $\sim O(1) N$, where $N$ is the length of the chain.
We would like to make a single occupancy of $\ket{D}$ costs more energy than that.
This motivates us to take $U\sim 100N$, where $100$ is a safety factor we somewhat randomly chose.
\item The constant term $1/4$ in the expression \eqref{U} of $\ket{D}\bra{D}$ does not affect the physics but should also be kept in the code. 
Otherwise, the ground state energy is dominated by terms of order $UN$, which confuses the DMRG solver.
\item It is better to choose the initial wavefunction to have no overlap with $\ket{D}\bra{D}$,
so \texttt{ramdomMPS(sites)} should better be avoided.
In our case we can simply use \texttt{MPS(InitState(sites,"Up"))}.
\end{itemize}
The \texttt{ITensor} code is given at the end of the Supplemental Material. 
It is a simple modification of a combination of the sample codes available on the \texttt{ITensor} website \cite{ITensor}.

\subsection{Low-lying spectrum of the converted 3-state Potts model}
Here we show that the low-lying spectrum of the converted 3-state Potts model is exactly equal to that of the original 3-state Potts model, for a sufficiently large $U$.
To see this, we note that the operator $\ket{D}_i\bra{D}_i$, for each $i$, commutes with the converted Hamiltonian $H$ \eqref{convertedH}.
Therefore we can simultaneously diagonalize $H$ together with all of $\ket{D}_i\bra{D}_i$.
The sector where $k$  of $\ket{D}_i$  ($i=1,\ldots,N$) are occupied has energy of order $Uk + O(1)N$. 
Now we take $U\gg N$.
Then, all states whose energy is of order at most $N$ have all $\ket{D}_i$ states unoccupied.
For these states the Hamiltonian \eqref{convertedH} reduces to the Hamiltonian of the original 3-state Potts model,
and therefore have exactly the same energy eigenvalues.

\subsection{The \texttt{ITensor} code}
\small
\begin{lstlisting}
#include "itensor/all.h"
using namespace itensor;

inline int mod(int x,int N){
    if(x>N)
        return x-N;
    return x;
}
int main(){       
    int N = 100;
    auto sites = SpinHalf(2*N,{"ConserveQNs=",false});
    //
    // Factors of 8, 4 and 2 are to rescale
    // spin operators into Pauli matrices
    //
    Real U=10000;
    Real f=1;
    Real J=1;
    auto ampo = AutoMPO(sites);
    for(int j = 1; j <= N; ++j){
        int i=2*j-1;
        ampo+=f*0.5*2,"Sz",mod(i+1,2*N);
        ampo+=-f*0.7071067811865475*4,"Sx",i,"Sx",mod(i+1,2*N);
        ampo+=f*0.7071067811865475*4,"Sy",i,"Sy",mod(i+1,2*N);
        ampo+=-f*0.5*4,"Sz",i,"Sz",mod(i+1,2*N);
        
        ampo+=-J*0.375*4,"Sx",mod(i+1,2*N),"Sx",mod(i+3,2*N);
        ampo+=J*0.375*8,"Sx",mod(i+1,2*N),"Sz",mod(i+2,2*N),"Sx",mod(i+3,2*N);
        ampo+=-J*0.125*4,"Sz",mod(i+1,2*N),"Sz",mod(i+3,2*N);
        ampo+=-J*0.25*4,"Sz",mod(i+1,2*N),"Sz",mod(i+2,2*N);
        ampo+=-J*0.125*8,"Sz",mod(i+1,2*N),"Sz",mod(i+2,2*N),"Sz",mod(i+3,2*N);
        ampo+=-J*0.25*4,"Sz",i,"Sz",mod(i+3,2*N);
        ampo+=-J*0.5*4,"Sz",i,"Sz",mod(i+2,2*N);
        ampo+=-J*0.25*8,"Sz",i,"Sz",mod(i+2,2*N),"Sz",mod(i+3,2*N);
        ampo+=J*0.375*8,"Sz",i,"Sx",mod(i+1,2*N),"Sx",mod(i+3,2*N);
        ampo+=-J*0.375*16,"Sz",i,"Sx",mod(i+1,2*N),"Sz",mod(i+2,2*N),"Sx",mod(i+3,2*N);
        ampo+=-J*0.125*8,"Sz",i,"Sz",mod(i+1,2*N),"Sz",mod(i+3,2*N);
        ampo+=-J*0.25*8,"Sz",i,"Sz",mod(i+1,2*N),"Sz",mod(i+2,2*N);
        ampo+=-J*0.125*16,"Sz",i,"Sz",mod(i+1,2*N),"Sz",mod(i+2,2*N),"Sz",mod(i+3,2*N);
        
        ampo+=-U*0.25*2,"Sz",mod(i+1,2*N);
        ampo+=U*0.25*2,"Sz",i;
        ampo+=-U*0.25*4,"Sz",i,"Sz",mod(i+1,2*N);
        ampo+=U*0.25*4,"Sz",i,"Sz",i;
    }
    auto H = toMPO(ampo);
    //
    // Set the parameters controlling the accuracy of the DMRG
    // calculation for each DMRG sweep. 
    //
    auto sweeps = Sweeps(60); 
    sweeps.maxdim() = 10,20,100,100,200,200,300,300,400;
    sweeps.cutoff() = 1E-10;
    sweeps.niter() = 2;
    sweeps.noise() = 1E-7,1E-8,0.0;
    //
    // Begin the DMRG calculation
    // for the ground state
    //
    auto [en,psi] = dmrg(H,MPS(InitState(sites,"Up")),sweeps,{"Quiet=",true});
    //
    // Compute the entanglement entropy
    //
    for(auto b=1;b<=2*N;b++){
        psi.position(b);
        //SVD this wavefunction to get the spectrum
        //of density-matrix eigenvalues
        auto l = leftLinkIndex(psi,b);
        auto s = siteIndex(psi,b);
        auto [U,S,V] = svd(psi(b),{l,s});
        auto u = commonIndex(U,S);
        //Apply von Neumann formula
        //to the squares of the singular values
        Real SvN = 0.;
        for(auto n : range1(dim(u)))
            {
            auto Sn = elt(S,n,n);
            auto p = sqr(Sn);
            if(p > 1E-12) SvN += -p*log(p);
            }
        printfln("{%d,  %.10f},",b,SvN);
    }
    return 0;
}
\end{lstlisting}

\end{document}